\documentclass[aps,prl,twocolumn]{revtex4}
\usepackage{epsf}
\newcommand{\Av}{\mbox{{\bf A}}}
\newcommand{\Rv}{\mbox{{\bf R}}}
\newcommand{\lv}{\mbox{{\bf l}}}

\begin{document}
%\widetext
 
\title{Upward curvature of the upper critical field\\
       in the Boson--Fermion model}

\author{Tadeusz\ Doma\'nski}
\affiliation{Centre de Recherches sur les Tr\`es Basses Temperatures
         CNRS, 38042 Grenoble, France \\
         and Institute of Physics, M.\ Curie Sk\l odowska University, 
         20-031 Lublin, Poland}
\author{Maciej M. Ma\'ska}
\author{Marcin Mierzejewski}
\affiliation{Institute of Physics, University of Silesia,
         40-007 Katowice, Poland}

\date{\today}

\begin{abstract}
We report on a non--conventional temperature behavior
of the upper critical field ($H_{c2}(T)$) which is found
for the Boson--Fermion (BF) model. We show that the BF model properly
reproduces two crucial features of the experimental data obtained
for high--$T_c$ superconductors: $H_{c2}(T)$
does not saturate at low temperatures and has an upward curvature.
Moreover, the calculated upper critical field fits very well the experimental
results. This agreement holds also for overdoped compounds, where a
purely bosonic approach is not applicable.
\end{abstract}

%74.20.-z, % Theories and models of superconducting state
%74.20.Mn, % Nonconventional mechanisms ...
%74.25.Bt, % Thermodynamic properties  
%71.10.-w  % Theories and models of many-electron systems
%}
  
\maketitle

%\section{Introduction}
%
Some of the unusual properties observed experimentally 
in the high temperature superconductors (HTSC) 
can be explained in terms of the effective two component 
models, which involve boson and fermion degrees of freedom. 
There are numerous examples of such theoretical scenarios,
for instance: electron fields coupled to gauge fluctuations, 
the RVB spinon-holon theory, the coupled electron-phonon 
systems, etc. In this paper we consider one of these 
models, where the conduction band particles (fermions) 
coexist and interact with the localized electron pairs
(hard-core bosons). This, so called {\em Boson Fermion} (BF) 
model, had initially been introduced \cite{Ranninger-95} 
a couple of years before the discovery of HTSC materials. 
The authors have conjectured that it might describe an 
emerging physics of the electron-phonon systems 
for the intermediate coupling strength.  

From a formal point of view, the Hamiltonian of this BF
model can be considered as an effective one for several 
models, which are often used in the solid state physics. 
It has been derived so far:  
(1) from the generalized periodic Anderson model with the 
large on-site attraction by eliminating the hybridization 
between the wide and narrow band electrons via canonical 
transformation \cite{Robaszkiewicz-87},
(2) from a purely fermionic single band system described by 
the extended Hubbard model using the concept of bosonization 
for lattice fermions \cite{Friedberg-94},
(3) from the Hubbard model on the two dimensional
plaquettized lattice in the strong interaction limit 
using the contractor method \cite{Auerbach-02}, and 
(4) from the resonating valence bond state of the $t-J$ 
model in the path integral technique  
\cite{Kochetov-02}.

The same BF model has been considered also by many other 
authors who postulated it in a more {\em ad hoc} way.
For example in the Ref.\ \cite{Enz-96} author has proposed
the BF type charge transfer Hamiltonian basing his arguments
on interpretation of the optical experiments. Other ideas
could be found in the Ref.\ \cite{Geshkenbein-97}. Authors
have represented the patches of the 2D Brillouin zone 
near the so called {\em hot spots} via dispersionless 
bosons, which are coupled to fermions of 
a remaining part of the Brillouin zone. The BF model is 
not only considered in a context of HTSC.
There are recent attempts to apply the same type of picture
for a description of the magnetically trapped atoms
of alkali metals \cite{Holland-01}.

Unconventional mechanism of superconductivity within
the BF model has been studied in a number of papers
\cite{Ranninger-85,Robaszkiewicz-87,Geshkenbein-97,
Friedberg-89,Ioffe-89,Micnas-90,Ranninger-95,Micnas-01}.
Fermions acquire superconducting coherence via exchange 
of the hard-core bosons. The same processes 
lead also to the effective mobility of bosons. 
Fermions/bosons can undergo a phase transition 
into the superconducting/superfluid phase at 
identically the same critical temperature 
$T_{c}=T_{sc}^{F}=T_{BE}^{B}$ \cite{Kostyrko-96}. 
It is worth to point out some of the unusual properties 
obtained for the BF model which are known in the most of
the HTSC materials: 
($i$) a non-BCS ratio $\Delta_{s}(T=0)/kT_{c} > 4$ 
    (except for the far under- and over-doping 
    regimes) \cite{Micnas-90,Ranninger-95};
($ii$) linear in $T$ resistivity of a normal phase 
    up to very high temperatures \cite{Eliashberg-87};
($iii$) change of sign of the Hall constant above $T_{c}$ and 
    the anomalous Seebeck coefficient \cite{Geshkenbein-97};
($iv$) appearance of the pseudogap in a normal phase 
    for temperatures $T^{*} > T > T_{c}$
    \cite{perturbative,DMFT,Domanski-01};
($v$) a particle-hole asymmetry of the single particle
    excitation spectrum in the normal phase 
    \cite{Geshkenbein-97,Domanski-02}. 

Some of unusual properties of HTSC are related to the
upper critical field. $H_{c2}$ can achieve values
of a few hundred Tesla. Moreover, 
the resistivity measurements for HTSC clearly show 
an upward curvature of $H_{c2}(T)$,  
with no evidence of saturation even at low temperatures
\cite{osofsky,mackenzie}. From the theoretical point of view the
upward curvature of the critical field occurs for instance in:
Bose-Einstein condensation of charged bosons \cite{alex}, Josephson tunneling
between superconducting clusters \cite{geshkenbein}, and in mean--field--type
theory of $H_{c2}$ with a strong spin--flip scattering \cite{kresin}.
However, this feature cannot be explained within
a conventional theory of $H_{c2}$ \cite{gorkov}. Therefore, it is a natural
test for theoretical approaches to HTSC. 
 
In the present study we show that the upward curvature of $H_{c2}(T)$  
is an intimate feature of the BF model. This result strongly supports 
the BF Hamiltonian as a model of HTSC.

%\section{The model}

We consider the Hamiltonian of the two--dimensional BF system,
immersed in a perpendicular, uniform magnetic field
\begin{eqnarray}
H^{BF} & = & \sum_{i,j,\sigma} \left( t_{ij}(\Av) -
\delta_{ij} \mu \right) c_{i\sigma}^{\dagger} c_{j\sigma} 
\nonumber \\
& + &  \sum_{i} \left( \Delta_{B}  - 2\mu \right) 
b_{i}^{\dagger} b_{i} 
+ 
v \sum_{i} \left(  b_{i}^{\dagger} c_{i\downarrow}
c_{i\uparrow} + {\rm h.c.} 
%b_{i} c_{i\uparrow}^{\dagger} c_{i\downarrow}^{\dagger} 
\right) \;.
\label{BF}
\end{eqnarray}
We use the standard notation for annihilation (creation) 
operators of fermion $c_{i\sigma}$ ($c_{i\sigma}^{\dagger}$) 
with spin $\sigma$ and of the hard core boson $b_{i}$ 
($b_{i}^{\dagger}$) at site $i$. Fermions interact
with bosons via the charge exchange interaction 
$v$. $\mu$ is the chemical potential and
 $t_{ij}(\Av)$ is
the hopping integral that depends on the magnetic field
through the vector potential $\Av$:
$$
t_{ij}\left(\Av \right)= t_{ij}(0)
\exp\left(\frac{ie}{\hbar c} \int^{\Rv_{i}}_{\Rv_{j}}
\Av\cdot d\lv\right).
$$  
To proceed, we apply the mean field decoupling 
for the boson fermion interaction
\begin{eqnarray}
b_{i}^{\dagger} c_{i\downarrow}c_{i\uparrow} 
\simeq \left< b_{i} \right>^{*}
c_{ i\downarrow} c_{i\uparrow} + 
b_{i}^{\dagger} \left< c_{ i\downarrow} 
c_{i\uparrow} \right>, 
\label{decoupling}
\end{eqnarray} 
which is justified when $v$ is small enough in 
comparison to the kinetic energy of fermions.
After the decoupling (\ref{decoupling}) we deal 
with the effective Hamiltonian composed of the  
fermion and boson contributions $H \simeq H^{F}+H^{B}$,
coupled through the selfconsistently determined expectational values
$\left< c_{i\downarrow} c_{i\uparrow} \right>$ and $\left< b_{i} \right>$
\cite{Robaszkiewicz-87,Ranninger-95}
\begin{eqnarray}
H^{F} & = &  \sum_{i,j,\sigma} \left[ t_{ij}(\Av)
-\delta_{ij} \mu \right]
c_{i\sigma}^{\dagger} c_{j\sigma} 
+ 
\sum_{i} \left(  \rho_{i}^{*} c_{i\downarrow}
c_{i\uparrow} +{\rm h.c.}  
%\rho_{i} c_{i\uparrow}^{\dagger}c_{i\downarrow}^{\dagger} 
\right), \nonumber \\
&&
\label{H_F} \\
H^{B} & =  & \sum_{i}\left [ \left( \Delta_{B}
-2\mu \right) b_{i}^{\dagger} b_{i} +
\Delta_i \; b_{i}^{\dagger} + \Delta_i^{*} b_{i} \right] \;,
\end{eqnarray}
where $\Delta_i=v\left< c_{i\downarrow} c_{i\uparrow} 
\right>$ and $\rho_{i}=v\left< b_{i} \right>$.  

One can exactly diagonalize the bosonic subsystem using a suitable 
unitary transformation. Statistical expectation values 
of the number operator $b_{i}^{\dagger}b_{i}$ and the 
parameter $\rho_{i}$ are given by 
\cite{Robaszkiewicz-87,Ranninger-95}
\begin{eqnarray}
\left< b_{i}^{\dagger} b_{i}\right> & = & 
\frac{1}{2} - \frac{\Delta_{B}-2\mu}{4\gamma_{i}} 
\tanh{\left(\frac{\gamma_{i}}{kT}\right)}, 
\label{nB} \\
\rho_{i} & = & - \; \frac{v\Delta_i}{2\gamma_{i}}\tanh{\left(
\frac{\gamma_{i}}{kT}\right)},
\label{rho}
\end{eqnarray}
where $\gamma_{i}=\frac{1}{2}\sqrt{(\Delta_{B}
-2\mu)^{2}+4|\Delta_i|^{2}}$ 
and $k$ is the Boltzmann constant.
At the phase transition
the superconducting order parameter
is infinitesimally small and one can expand
$\rho_i$ in powers of $\left< c_{i\downarrow} c_{i\uparrow} 
\right> $ up to the leading order. Then, the fermionic
subsystem is described by
\begin{eqnarray}
H^{F} & = &  \sum_{i,j,\sigma} \left[ t_{ij}(\Av)
-\delta_{ij} \mu \right]
c_{i\sigma}^{\dagger} c_{j\sigma} \nonumber \\
&-& 
V(T)\sum_{i} \left(\left<c^{\dagger}_{i\uparrow}
c^{\dagger}_{i \downarrow} \right> c_{i\downarrow}
c_{i\uparrow} +{\rm h.c.}  
%\rho_{i} c_{i\uparrow}^{\dagger}c_{i\downarrow}^{\dagger} 
\right), \label{H_MF}
\end{eqnarray} 
where \begin{equation}
V(T)=\frac{v^{2}}{\Delta_{B}-2\mu} 
\tanh\left(\frac{\Delta_{B}-2\mu}{2kT} \right).
\label{V_MF}
\end{equation}

The effective Hamiltonian (\ref{H_MF})
represents a BCS--type model with
isotropic on--site pairing. 
The boson--fermion coupling enters $H^{F}$
through the pairing interaction
(\ref{V_MF}) and the chemical potential.
The latter quantity should be evaluated
from the conservation of the total charge
$n_{\rm tot}\equiv 2n_B+n_F=
2/N \sum_i \left<b^{\dagger}_i b_i\right>
+ 1/N \sum_{i \sigma} 
\left<c^{\dagger}_{i \sigma}  c_{i \sigma}\right>$.
We restrict further investigation only
the nearest--neighbor hoping, when 
the fermionic energy spectrum
is known as the Hofstadter butterfly \cite{hofstadter}. 

First, we assume that the bosonic level is in the middle of 
the fermionic band ($\Delta_B=0$) and
$n_{\rm tot}=2$.  It is the simplest case, when the parameters
of the Hamiltonian (\ref{H_MF}) can be evaluated
without a numerical investigation of the Hofstadter
butterfly. Namely, since the Hofstadter energy spectrum 
is symmetric, for $\mu=0$ one gets
$n_F=1$, $n_B=1/2$, $n_{\rm tot}=2$ 
and $V(T)=v^2/(2kT)$. 
In order to calculate $H_{c2}$ one can apply
a two--dimensional version
of the Helfand-Werthamer (HW) theory \cite{kresin}, where 
the coupling constant, $\lambda$, depends
on temperature. Since, $\lambda= V \rho_{\rm FS}$, where
$\rho_{\rm FS}$ is the density of states at the Fermi level,
one gets
\begin{equation}\lambda(T)= \lambda(T_c) \frac{T_c}{T}.
\label{lambda}
\end{equation} 
The HW theory was derived
for a free--electron gas and 
neglects the Landau level structure
(so called quasiclassical limit).   
However, for a weak magnetic field 
these effects do not lead to an essential modification
of $H_{c2}(T)$ \cite{nasz}.
Figure 1 shows $H_{c2}(T)$
calculated within the HW approach with the coupling constant
determined by Eq. (\ref{lambda}).  
%There are two crucial
%facts which qualitatively determine the critical field: 
In order to predict the qualitative 
temperature dependence of the critical field
it is sufficient to recall the following facts: 
($i$) when $\lambda(T)=\rm const$ (standard HW theory), 
$H_{c2}(T)$ is linear for a weak magnetic field;
($ii$) in the BF model $\lambda$ increases with the
decrease of temperature and
diverges when $T \rightarrow 0$ (Eg. \ref{lambda}). Then,
it becomes obvious that BF model properly
reproduces two crucial features of the experimental data:
$H_{c2}(T)$ does not saturate at low temperatures
and has an upward curvature, at least for a weak
magnetic field. Numerical results (see Fig. 1)
show that the latter property holds in the whole
range of temperature.
\begin{figure}
\epsfxsize=6.5cm
\centerline{\epsffile{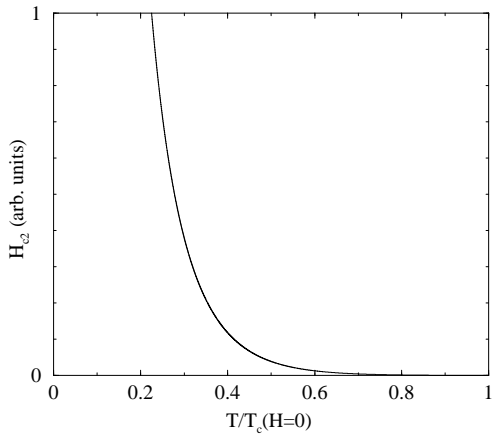}}
\caption{Temperature dependence of the upper critical field,
obtained from the HW approach with the coupling constant given by 
Eq.(\ref{lambda}) . We have assumed $kT_c(H=0)=0.02t$.} 
\end{figure}

Next, we show that the BF model properly describes $H_{c2}(T)$ 
for a wide range of model parameters, when the problem cannot
be reduced to the effective HW theory. In this case we apply a lattice version
of the Gor'kov equations \cite{nasz}
\begin{equation}
\Delta_{i}
%\left< c_{i\downarrow} c_{i\uparrow} \right>
= \frac{V(T)}{\beta }\sum_{j,\omega_n}
%\left< c_{j\downarrow} c_{j\uparrow} \right>
\Delta_{j} 
G(i,j,\omega_n)G(i,j,-\omega_n).
\label{gorkov}
\end{equation}
Here, $G(i,j,\omega_n)$ is the one--electron Green's function
in the presence of a uniform and static magnetic field and
$\omega_n$ is the fermionic Matsubara frequency.
With the help of the Hofstadter approach \cite{hofstadter}, 
equation (\ref{gorkov}) can exactly be solved for clusters
of the order of $10^4$ lattice sites. For the details we refer to
Ref. \cite{nasz}. In contradistinction to the quasiclassical 
approaches (e.g., HW or Ginsburg--Landau theory) we explicitly
account for the actual structure of Landau levels. 

Figure 2 shows the upper critical field obtained for
$n_{\rm tot}=1$ and
different positions of the bosonic level.
We consider two cases: ($i$)
when this level is below the Fermi
energy ($\Delta_B<0$) there is a finite 
number of bosons also at $T\rightarrow 0$, 
($ii$) for $\Delta_B >0$ bosonic states are occupied 
only virtually.
\begin{figure}
\epsfxsize=6.5cm
\centerline{\epsffile{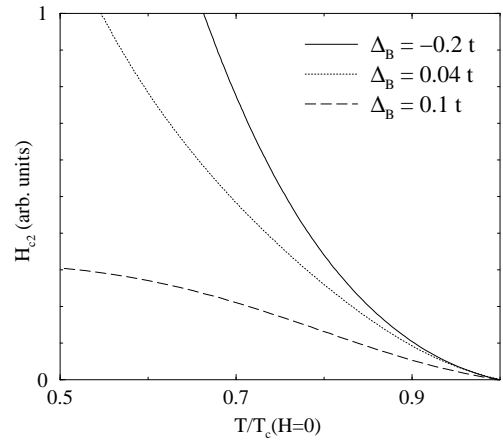}}
\caption{$H_{c2}(T)$ obtained from Eq. (\ref{gorkov}) for $n_{\rm tot}=1$
and different bosonic levels. The fermion--boson coupling was adjusted
to obtain the same critical temperature in the absence of magnetic field.}
\end{figure}
The upward curvature of $H_{c2}(T)$ appears predominantly
in the first case (solid line in Fig.2). 
When $\Delta_B$ is shifted above the Fermi energy,
the curvature is  gradually reduced.
Finally, when $\Delta_B \gg kT $   
the curvature changes from
positive to negative (dashed line in Fig. 2) and one
reproduces standard results for a purely fermionic
system \cite{nasz}. In Fig. 2 we have not presented
results in the low temperature regime.
Although, we have carried out numerical calculations
for large clusters, this approach is not applicable at
genuinely low temperatures. In this case
the Cooper pair susceptibility
accounts only for very few fermionic states
(with energy close to the Fermi level)
instead of a continuous density of states. However,
one can prove that BF model qualitatively reproduces
$H_{c2}(T)$ also for $T\rightarrow 0$. 
Combining Eqs. (\ref{V_MF}) and (\ref{nB}) one can
express the effective pairing potential in terms of the
bosonic occupation number
\begin{equation}
V(T)=\frac{2 v^{2}}{\Delta_{B}-2\mu} 
\left(\frac{1}{2}-\left<b^{\dagger}_i b_i\right> \right).
\label{V_MF1}
\end{equation}   
It is straightforward to note that for 
$\left<b^{\dagger}_i b_i\right>=1/2$ one gets 
$\Delta_{B}=2\mu $ and Eq. (\ref{V_MF}) is reduced to $V(T)= v^2/2kT$. 
On the other hand, when $\left<b^{\dagger}_i b_i\right>\ne 1/2$ but
$0<n_B<1$ the denominator in Eq. (\ref{V_MF1}) 
vanishes for $T\rightarrow 0$. Therefore, 
$V(T)$ diverges for $T\rightarrow 0$ provided that
$0<n_b<1$. It means that a requirement of partial occupation of
bosonic states is sufficient to reproduce the experimental   
low temperature behavior of $H_{c2}$.
  
To complete the discussion, 
we show that BF model accurately reproduces the
experimental data (see Fig. 3).
\begin{figure}
\epsfxsize=6cm
\centerline{\epsffile{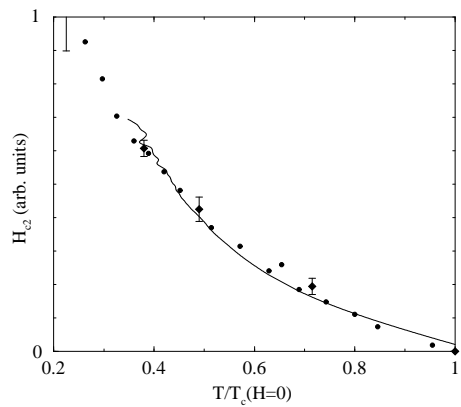}}
\caption{Fit to experimental results (continuous curve). The experimental data 
(points) were taken for ${\rm Tl}_2{\rm Ba}_2{\rm Cu0}_6$ from
Ref. \cite{mackenzie}, and for 
${\rm Bi}_2{\rm Sr}_2{\rm CuO}_y$ from Ref. \cite{osofsky}.}
\end{figure}
We have chosen an appropriate set of model parameters, 
for which $H_{c2}(T)$ fits very well
the results presented in Refs. \cite{osofsky,mackenzie}.
The theoretical curve was calculated for bosonic
level that is situated slightly above the Fermi
energy ($n_{\rm tot}=1$ and $\Delta_B=0$).
In this case $n_F \gg n_B$ and the BF system is mainly of fermionic
character. The experimental data were obtained for overdoped
compounds, when HTSC exhibits a Fermi liquid character \cite{condmat}.
Therefore, BF model is fully consistent with fermionic--type 
behavior of the overdoped HTSC, whereas the purely Bosonic 
approach \cite{alex} is not applicable in this regime.
 
To conclude, we have shown that the BF model accurately  
describes the upper critical field observed in HTSC.
It is known that the BF model correctly describes
the pseudogap phenomenon and some other unconventional 
normal--state properties of HTSC.  
In this letter we have
shown that this model reproduces also other unusual feature
of HTSC, which cannot be explained within a standard
BCS--type approach. 
Moreover, our results remain in agreement with the 
Fermi liquid behavior of the overdoped compounds. 
Therefore, it is an important support for the  
BF Hamiltonian as a model of HTSC.

%\end{multicols}

\begin{thebibliography}{11}
\bibitem{Ranninger-85}
    J.~Ranninger and S.~Robaszkiewicz, Physica 
    {\bf B 135}, 468 (1985).
\bibitem{Robaszkiewicz-87}
    S.~Robaszkiewicz, R.~Micnas and J.~Rannin\-ger, 
    Phys.\ Rev.\ {\bf B 36}, 180 (1987).
\bibitem{Friedberg-94}
    R.~Friedberg, T.D.~Lee and H.C.~Ren, Phys.~Rev.\
    {\bf B 50}, 10190 (1994).
\bibitem{Auerbach-02}
    E.~Altman and A.~Auerbach, Phys.~Rev.\ {\bf B 65},
    104508 (2002).
\bibitem{Kochetov-02}
    E.~Kochetov and M.~Mierzejewski, cond-mat/0204420.
\bibitem{Enz-96}
    Ch.P.~Enz, Phys.~Rev.\ {\bf B 54}, 3589 (1996).
\bibitem{Geshkenbein-97}
    V.B.~Geshkenbein, L.B.~Ioffe and A.I. Larkin, Phys.~Rev.\
    {\bf B 55}, 3173 (1997).
\bibitem{Holland-01}
    M.~Holland, S.J.J.M.F.~Kokkelmans, M.L.~Chiofalo
    and R.~Walser, Phys.~Rev.~Lett.\ {\bf 87}, 120406 (2001);
    E.~Timmermanns et al, Phys.~Let.~A {\bf 285}, 228 (2001);
    Y.~Ohashi and A.~Griffin, cond-mat/0201262 (preprint).
\bibitem{Eliashberg-87}
    G.M.~Eliashberg, Pis'ma Zh.~Eksp.~Teor.~Fiz.\ {\bf 46}, 
    94 (1987).    
\bibitem{Friedberg-89}
    R.~Friedberg and T.D.~Lee, Phys.~Rev.\  {\bf B 40}, 423 (1989);
    R.~Friedberg, T.D.~Lee and H.C.~Ren, Phys.~Lett.~A {\bf 152},
    417 (1991).
\bibitem{Ioffe-89}
    L.~Ioffe, A.I.~Larkin, Y.N.~Ovchinnikov and L.~Yu, 
    Int.~Journ.~Modern~Phys.\ {\bf B 3}, 2065 (1989).    
\bibitem{Micnas-90}
    R.~Micnas, J.~Ranninger and S.~Ro\-basz\-kie\-wicz,
    Rev.\ Mod.\ Phys.\ {\bf 62}, 113 (1990).
\bibitem{Ranninger-95}    
    J.~Ranninger and J.M.~Robin, Physica {\bf C 253}, 279 (1995). 
\bibitem{Micnas-01}
    R.~Micnas, S.~Robaszkiewicz and B.~Tobijaszewska, 
    Physica {\bf B 312}-{\bf 313}, 49 (2002);
    R.~Micnas and B.~Tobijaszewska, Acta Phys. Pol.\
    {\bf B 32}, 3233 (2001).
\bibitem{Kostyrko-96}
    T.~Kostyrko and J.~Ranninger, Phys.~Rev.\ {bf B 54},
    3241 (1996).
\bibitem{perturbative}
    J.~Ranninger, J.M.~Robin, M.~Eschrig, Phys.~Rev.~Lett.\ 
    {\bf 74}, 4027 (1995);
    J.~Ranninger and J.M.~Robin, Solid State Commun.\
    {\bf 98}, 559 (1996);
    J.~Ranninger and J.M.~Robin, Phys.~Rev.\ {\bf B 53}, 
    R11961 (1996);
    H.C.~Ren, Physica {\bf C 303}, 115 (1998);    
    P.~Devillard and J.~Ranninger, Phys.~Rev.~Lett.\ {\bf 84}, 
    5200 (2000).
\bibitem{DMFT}
    J.M.~Robin, A.~Romano, J.~Ranninger, Phys.~Rev.~Lett.\
    {\bf 81}, 2755 (1998); 
    A.~Romano and J.~Ranninger, Phys.~Rev.\ {\bf B 62}, 4066 (2000).
\bibitem{Domanski-01}
    T.~Doma\'nski and J.~Ranninger, Phys.~Rev.\ {\bf B 63}, 
    134505 (2001).
\bibitem{Domanski-02}
    T.~Doma\'nski and J.~Ranninger, (2002) {\em unpublished}.
\bibitem{osofsky} 
    M.S.~Osofsky {\it et al}, Phys. Rev. Lett. {\bf 71},
    2315 (1993).
\bibitem{mackenzie} 
    A.P.~Mackenzie {\it et al}, Phys. Rev. Lett. {\bf71},  
    1238 (1993). 
\bibitem{alex} 
    A.S.~Alexandrov {\it et al}, Phys. Rev. Lett. 
    {\bf 76}, 983 (1996).
\bibitem{geshkenbein} 
    V.B.~Geshkenbein,L.B.~Ioffe, and A.J.Millis, 
    Phys. Rev. Lett. {\bf 80}, 5778 (1998).  
\bibitem{kresin} 
    Yu.N.~Ovchinnikov and V.Z.~Kresin, Phys. Rev. {\bf B 52}, 
    3075 (1995).  
\bibitem{gorkov}  
    L.P. Gor'kov, Zh. Eksp, Teor. Fiz. {\bf36}, 1918 (1959)
    [Sov. Phys. JETP {\bf9}, 1364 (1960)].
\bibitem{hofstadter} 
    R.D. Hofstadter, Phys. Rev. {\bf B 14}, 2239 (1976).
\bibitem{nasz} 
    M. M. Ma{\'s}ka and M. Mierzejewski, Phys. Rev. {\bf B 64}, 
    064501 (2001).
\bibitem{condmat}  
   Y. Kubo {\it et al.}, Phys. Rev. {\bf B 43}, 7875 (1991);
   M. Suzuki and M. Hikita, Phys. Rev. {\bf B 44}, 249 (1991);
    C. Proust, E. Boakin, R. W. Hill, L. Taillefer, and
A. P. Mackenzie, {\tt cond-mat/0202101}.
\end{thebibliography}
\end{document}